 \documentclass[final,5p,times,twocolumn,authoryear]{elsarticle}

\usepackage{amssymb}
\usepackage{lipsum}
\usepackage{amsmath}	
\usepackage[colorlinks=true,linkcolor=blue,citecolor=blue,urlcolor=blue]{hyperref}

\newcommand{\pcm}{\textrm{pc\,cm}$^{-3}$}
\journal{New Astronomy}

\begin{document}

\begin{frontmatter}



\title{Disentangling propagation effects from Fast Radio Burst spectra: An analysis on simulated data
}


\author[first]{Aishwarya Kumar}
\author[second]{Fereshteh Rajabi}
\author[first]{Martin Houde}
\affiliation[first]{organization={Department of Physics and Astronomy,\ The University of Western Ontario },
            addressline={ 1151 Richmond Street}, 
            city={London},
            postcode={N6A 3K7}, 
            state={Ontario},
            country={Canada}}
\affiliation[second]{organization={Department of Physics and Astronomy, \ McMaster University},
            addressline={ 1280 Main Street West}, 
            city={Hamilton},
            postcode={L8S 4L8}, 
            state={Ontario},
            country={Canada}}

\begin{abstract}
We present a methodology to decouple propagation effects, specifically scattering and dispersion, from the intrinsic spectro-temporal properties of repeating Fast Radio Bursts. Utilizing the Triggered Relativistic Dynamical Model, and assuming superradiance as the emission mechanism, we generate simulated sub-bursts and inject controlled levels of scattering and residual dispersion. For each burst, we measure the sub-burst slope, defined as the trajectory of the centroids in the dynamic spectrum, and the characteristic duration of the burst profile. We then fit a modified sub-burst slope law to the resulting slope-duration measurements to recover the scattering timescale, residual dispersion measure, and other model parameters. Under the thin-screen approximation, the scattering timescale at $1~\mathrm{GHz}$ is precisely recovered, typically to within $\sim 1-2\%$ of the true value. In contrast, the residual dispersion is recovered with comparatively lower precision, with median absolute errors of $\sim 0.3-0.6$~\pcm, reflecting its weaker constraint and degeneracy with an intrinsic parameter. Despite this, the modified sub-burst slope law successfully reproduces the spectro-temporal evolution and accurately constrains the scattering properties even for diverse intrinsic burst populations. These results demonstrate that our framework yields a tractable method for separating propagation-induced distortions from intrinsic emission characteristics to a meaningful degree, enabling more reliable inference of the physical properties of FRB sources.

\end{abstract}




\end{frontmatter}




\section{Introduction}
\label{introduction}

Fast radio bursts are a form of highly energetic radio transients of sub-second duration \citep{Lorimer_2007, Review_ppr_2019, cordes2019fast}. Observations have revealed distinct characteristics that differentiate these events from other radio-frequency transients, such as Galactic pulsars. FRBs are broadly classified as either repeating or non-repeating. These two categories appear to have distinct burst morphologies and properties, potentially indicating different progenitor sources \citep{Hashimoto_2020, Chime_2021, Pleunis_2021, Zhang_2023}. Particularly, repeating FRBs display remarkably complex time-frequency structures that are intrinsic to the bursts.  A temporally resolved dynamic spectrum of such bursts may reveal either a single, broadband pulse or multiple, narrowly confined components known as sub-bursts. Two key observables describe their spectral structure: the drift rate and the sub-burst slope. Drift rate is the downward drift in frequency between different resolved components of a burst. Individual components of a burst, or sub-bursts, also show this downward drift in frequency, which constitutes the sub-burst slope. However, rare exceptions exhibiting upward frequency drift, sometimes referred to as ``happy trombone'' bursts, have also been reported in $\mathrm{FRB\ 20201230B}$ \citep{Faber_2024}. The spectro-temporal features extracted from the burst morphology can be utilized to inform the nature of the emission process. 

Based on their energies, timescales, repeatability, and other observational features, there are a range of progenitor models proposed that usually involve a compact object. One of them is the Triggered Relativistic Dynamical Model (TRDM) proposed by \citet{rajabi2020simple}, which we use in our analysis. The model predicts an inverse relationship between the sub-burst slope and duration, an inverse scaling of duration with frequency, and a quadratic dependence of the sub-burst slope on frequency. These predictions have been substantiated across 10 sources (with over 1000 bursts) spanning frequencies from 149 to 7500 MHz by \citet{rajabi2020simple, chamma2021evidence, Wang_2022, chamma2023broad, jahns2023frb, brown2023validating, Chamma_2025}. Nevertheless, some sub-bursts deviate from these trends, particularly at lower frequencies, suggesting the presence of a frequency dependent propagation or emission process. One such process capable of producing these spectral distortions in the dynamic spectrum is scattering. The extragalactic nature of FRBs exposes them to the plasma inhomogeneities in the intervening galactic and inter-galactic regions which introduces a frequency dependent variation in the burst profile. As a consequence, even an intrinsically narrow burst would appear to be temporally smeared. This asymmetrical broadening of the burst profile can skew the relationships between the observed parameters. Another important propagation effect is imperfect de-dispersion, which introduces residual dispersive delays that modify the burst structure, particularly the measured sub-burst slope.

\citet{Kumar_2026} (hereafter, Paper~I) established a theoretical framework to assess the impact of the propagation effects on the sub-burst slope law, demonstrating that scattering and dispersion lead to clear deviations from the ideal relation predicted by the TRDM. In the present work, we test this framework using simulated bursts to assess its ability to recover physical parameters under controlled conditions. In a forthcoming study (Paper~III), we will apply the algorithm to a sample of real repeating FRB data. Given the limited availability of observations spanning broad frequency ranges, simulations provide an essential proof-of-concept prior to application to real data. 

Our paper is structured as follows. In Section~\ref{sec:sm}, we describe the emission model used to simulate FRB-like bursts and the procedure to introduce scattering and dispersion. In Section~\ref{sec:meth}, we detail the construction of the simulated sub-bursts and the injection of dispersion and scattering effects. We also summarize the measurement pipeline used to extract sub-burst slopes and durations from the simulated data, and present the framework for simultaneously fitting these observables using the modified sub-burst slope law. Section~\ref{sec:res} discusses the recovery of key parameters, such as the residual dispersion, scattering timescales, and intrinsic burst properties, and assesses the robustness of the methodology. The key insights and results of our analysis are summarized in Section~\ref{sec:sum}.

\section{Theoretical Framework}
\label{sec:sm}
\subsection{Superradiance model of emission}
\label{sec:super} 
We employ the semi-classical model of superradiance (SR) to generate pulses that match the timescales and morphology of typical FRB emissions. SR is a coherent collective spontaneous emission process observed in a macroscopic ensemble of $N$ atoms (or molecules; but we use the former for our discussion) initially having a population inversion between the two levels of a given spectral transition. In such an ensemble, interactions through a common radiation field can establish correlations between the atomic dipoles. The system then behaves as a single entity emitting radiation coherently via enhanced cooperative spontaneous decay \citep{Dicke_1954}. The intensity of emission is proportional to $N^2$, in contrast to the spontaneous emission emanating from a group of independent atoms, where the intensity is proportional to $N$. Furthermore, it is found that the characteristic timescale of SR emission, $T_R$, scales as $1/N$ times the spontaneous emission timescale ($\tau_{\textrm{sp}}$) of the single-atom transition.

While SR has been an intense subject of research in the quantum optics community for more than fifty years \citep{Gross_Haroche_1982, Benedict_1996}, \citet{rajabi2016dicke} laid down the necessary framework for applying SR to astrophysical domains, such as for explaining powerful flaring events in maser-hosting regions \citep{ rajabi_2016_2, rajabi_2019, Rajabi_2020_maser, Rashidi_2025, Rashidi_2026}. The dynamics of SR emission is commonly modeled through the semi-classical Maxwell-Bloch equations that trace the evolution of the population inversion (i.e., the difference between the population densities of the upper and lower energy levels of the transition), polarization and electric field in long cylindrical samples \citep{rajabi2016dicke}. \citet{rajabi2020astronomical} describes the conditions under which a sample of inverted atoms or molecules in a masing region can undergo SR. Briefly, SR occurs when (\textit{i}) the column density of the population inversion exceeds a critical threshold  (\textit{ii}) the timescale of relaxation ($T_1$) and dephasing ($T_2$), for example related to collisions, are longer than the characteristic timescale of SR emission ($T_R$), and (\textit{iii}) atoms participating in the process should sustain sufficient velocity coherence to interact cooperatively (as is required for the masing action).

For regions harboring population inversion below the critical threshold, SR could still ensue with the help of a triggering pulse. This triggering pulse can either (i) increase the population inversion of the sample (e.g., via a ``pump'' acting at a shorter wavelength; \citealt{Gray_2012}) or, (ii) inject coherence in the system at the frequency of the transition to effectively lower the critical column density level. The sample needs some time to build up the coherence. Thus, we define the delay time, $\tau^\prime_\mathrm{D}$, as the time between the injection of the trigger at the input of the sample and the onset of SR pulse at the output minus the propagation time within it (i.e., $\tau^\prime_\mathrm{D}$ is a retarded-time interval in the source's reference frame). For FRBs, the triggering source could be a young and energetic radio pulsar, a magnetar or even a persistent radio source. \citet{houde2019triggered} developed and used this framework to model bursts emanating from FRB121102A. 

\subsection{Triggered Relativistic Dynamical Model (TRDM)}
\label{sec:trdm}
In this model, we have a triggering source and an emitting source, i.e., the FRB source, aligned along the line of sight of the observer. The emitting source is moving relativistically either towards or away from the observer. The motion of particles within the source's reference frame is assumed to be mildly relativistic with some level of velocity coherence. The arrival of the triggering pulse induces favorable conditions for superradiant emission, and after some delay $\tau^\prime_\mathrm{D}$, the source radiates an intense burst of coherent radio emission (see Figure 1 and 2 of \citealt{rajabi2020simple} for a schematic and temporal sequence of the model). We adopt the convention that primed quantities represent source-frame values, while unprimed quantities denote observer-frame measurements. Although SR is used here as the emission mechanism of the source, the TRDM formalism is more general and may also apply to other coherent emission processes subject to the same relativistic dynamics.

Mildly relativistic motions within the emitting region produce frequency-dependent temporal variations across a single sub-burst, transforming an intrinsically narrowband emission process into the broad observed spectro-temporal structure of FRBs. Under these assumptions, \citet{rajabi2020simple} showed that the frequency normalized sub-burst slope and duration obey the relation
\begin{equation}
    \frac{1}{\nu_{\textrm{obs}}}\frac{\mathrm{d} \nu_{\textrm{obs}} }{\mathrm{d} t_\textrm{D}}= -\frac{A}{t_\textrm{w}},
	\label{eq:sub_burst}
\end{equation}
where $\nu_{\textrm{obs}}$, $t_\textrm{D}$, $t_\textrm{w}$ are the frequency, delay and duration of the pulse in the observer's frame. This expression constitutes the sub-burst slope law. The parameter 
\begin{equation}
    A = \frac{\tau'_\textrm{w}}{\tau'_\textrm{D}} = \frac{t_\mathrm{w}}{t_\mathrm{D}}
	\label{eq:a_value}
\end{equation}
is a systemic quantity defined as the ratio of the intrinsic duration to the intrinsic delay. Fits of this law to observational data have been performed for several repeating FRB sources \citep{chamma2021evidence, chamma2023broad, brown2023validating, Chamma_2025}, yielding values of $A$ typically in the range of $0.08-0.3$. The intrinsic pulse is subjected to relativistic transformations, such that the delay and duration measured in the observer’s frame are related to the corresponding intrinsic source-frame quantities by
\begin{align}    
   t_\mathrm{D} & =\tau_\mathrm{D}'\frac{\nu^\prime}{\nu_{\mathrm{obs}}}
   \label{eq:td}\\
   t_\mathrm{w} & =\tau_\mathrm{w}'\frac{\nu^\prime}{\nu_{\mathrm{obs}}}
   \label{eq:tw}
\end{align}
where $\nu'$ denotes the emission frequency. Here, the observed frequency, $\nu_{\mathrm{obs}}$, is related to the source-frame emission frequency, $\nu'$, through the relativistic Doppler factor,
\begin{equation}
\nu_{\mathrm{obs}} = \nu' \sqrt{\frac{1+\beta}{1-\beta}}.
\label{eq:velspread}
\end{equation}
where $\beta$ is the velocity of the emitting region relative to the observer in units of the speed of light \citep{rajabi2020simple}. 

Equation (\ref{eq:tw}) therefore implies that bursts detected at higher frequencies would appear shorter in duration. Several observational studies have demonstrated this duration-frequency dependence \citep{chamma2023broad, brown2023validating, Chamma_2025}. These works report a characteristic scaling parameter of $\tau_\mathrm{w}'\nu^\prime \equiv t_0 \approx 1500\ \mathrm{ms\,MHz}$, albeit with some statistical scatter. In addition, \citet{Chamma_2025} suggested that ultra-FRBs \citep{Nimmo_2021, Snelders_2023} may follow a different scaling, potentially forming a distinct population in the $t_\mathrm{w}$–$\nu$ parameter space. 

Observational studies \citep{houde2019triggered, rajabi2020simple, chamma2023broad, brown2023validating} have further shown that the bandwidth of individual sub-bursts scales approximately linearly with their central frequency, such that
\begin{equation}
    \Delta \nu _{\mathrm{obs}} \simeq B_\nu\nu _{\mathrm{obs}}
	\label{eq:bw_1}
\end{equation}
with $\nu_{\mathrm{obs}}$ located at the central frequency of the sub-burst and $B_{\nu}$, an empirical proportionality constant, typically in the range $0.14-0.16$. Taken together, these relations form the basis for constructing wideband emission via Doppler broadening of an intrinsically narrow SR pulse, which for this study is centered at $1420$~MHz in the source frame.

\subsection{Scattering }
\label{sec:scat}
Multipath propagation of radio waves through inhomogeneous plasma causes asymmetrical temporal broadening of pulse profile. The observed signal can therefore be described as the convolution of the intrinsic pulse with the impulse response of the scattering medium. To model this effect, we adopt the thin-screen approximation, in which the scattering material is assumed to be confined to a localized region along the line of sight. Under this approximation, the pulse broadening function is represented by a one-sided exponentially decaying profile \citep*{Cronyn1970, Rickett1977}
\begin{equation}
    S(\nu, t)= \frac{1}{\tau_\mathrm{sc}}\mathrm{exp}\left ( -\frac{t}{\tau_\mathrm{sc}} \right )H(t),
	\label{eq:scat_1}
\end{equation}
where $H(t)$ is the Heaviside step function. The power law dependence of the scattering time-scale $\tau_\mathrm{sc}$ on the observing frequency is given by
\begin{align}
   \tau_\mathrm{sc} = \Lambda_\mathrm{sc}\left(\frac{\nu}{1\,\mathrm{GHz}}\right)^{-n}, 
   \label{eq:tau_1}
\end{align}
where, $n$ is the scattering index and $\Lambda_\mathrm{sc}$ is the scattering constant whose value depends on the fluctuations in electron density, the scale size of the scattering medium, and the distance to the object \citep{lorimer2005handbook}. Equivalently, it is defined as the scattering timescale evaluated at an observing frequency of 1 GHz. 

The observed pulse, $I(\nu,t)$, is generated through the convolution ($\ast$) of the superradiant source profile, $I_0(\nu,t)$, with the scattering function
\begin{equation}
    I(\nu,t)= I_0(\nu, t)\ast S(\nu, t).
	\label{eq:conv_1}
\end{equation}

As developed in Paper I, we approximate the intrinsic FRB signal as a one-sided exponential,
\begin{align}
    I_0\left(\nu, t\right) &= \frac{F_0}{t_\mathrm{w}}  \exp \left[ \frac{-(t - t_\mathrm{D})}{t_\mathrm{w}} \right] H(t),
	\label{eq:sr_int}
\end{align}
where $t_\mathrm{D}$ and $t_\mathrm{w}$ are the delay and duration as mentioned before, To ease the notation, we omit the Heaviside function from here on forward. Although exact FRB and SR pulse profiles can exhibit more complex temporal structure (see Fig.~\ref{fig:sr_profile}), we adopt a one-sided exponential approximation for its simplicity and analytical tractability. However, our methodology is applicable, in principle, to any intensity profile. The convolution of this profile with scattering is
\begin{align}
    I\left(\nu, t\right) &= \frac{F_0}{\tau_\mathrm{sc} - t_\mathrm{w}} \left\{ \exp \left[ \frac{-(t - t_\mathrm{D})}{\tau_\mathrm{sc}} \right] - \exp \left[ \frac{-(t - t_\mathrm{D})}{t_\mathrm{w}} \right] \right\}.
	\label{eq:con_fin}
\end{align}
For the condition where $\tau_\mathrm{sc} \approx t_\mathrm{w}$, we use the L'Hôpital's rule to find the limiting solution
\begin{align}
    \lim_{\tau_\mathrm{sc} \to t_\mathrm{w}}I\left(\nu, t\right) = F_0 \frac{(t-t_\mathrm{D})}{t_\mathrm{w}^2} \exp \left[\frac{-(t-t_\mathrm{D})}{t_\mathrm{w}}\right].
    \label{eq:con_lim}
\end{align}
Equations (\ref{eq:con_fin}) and (\ref{eq:con_lim}) are used to fit our generated sub-burst profiles. Although scintillation effects are present in observed bursts, for simplicity we have not included them in our analysis.

\subsection{ Residual dispersion }
\label{sec:disp}
Dispersion is typically recovered with appreciable precision using both coherent and incoherent dedispersion techniques \citep{petroff2019fast}. However, some residual dispersion can persist and introduce bias in measured burst properties, such as the sub-burst slope. We define this residual dispersion, $\Delta \mathrm{DM}$, following the convention introduced in Paper~I as 
\begin{align}
    \Delta\mathrm{DM}= \mathrm{DM}_\mathrm{true}-\mathrm{DM}_\mathrm{est},
    \label{eq:del_dm}
\end{align}
where $\mathrm{DM}_\mathrm{true}$ is the true DM of the source and $\mathrm{DM}_\mathrm{est}$ is the value estimated from observational techniques such as structure-maximizing DM, or signal-to-noise maximizing dedispersion. Therefore, under our convention, $\Delta \mathrm{DM}<0$ corresponds to over-dedispersion (over-correction), whereas $\Delta \mathrm{DM}>0$ corresponds to under-dedispersion (under-correction). 

The resulting dispersive delay present across the sub-burst bandwidth is given by
\begin{align}
   \Delta t_{\mathrm{DM}}= a \,\Delta\mathrm{DM} \left (\frac{1}{\nu^2} -\frac{1}{\nu^2_\mathrm{ref}} \right ) \: \: \mathrm{ms},
	\label{eq:dm_1}
\end{align}
where $a=4.148\,806\,4239(11)\;\textrm{GHz}^2\,\textrm{cm}^3\,\textrm{pc}^{-1}\,\textrm{ms}$ \citep{Kulkarni2020} and $\nu_{\textrm{ref}}$ is the reference frequency. In practice, $\nu_{\textrm{ref}}$ can be set to the highest frequency present in a dynamic spectrum or to infinity. In our model, this residual delay is incorporated by adding the $ \Delta t_{\mathrm{DM}}$ term to relativistic delay $t_\mathrm{D}$ in Equation (\ref{eq:con_fin}).

\section{Methodology}
\label{sec:meth}

\subsection{Simulating superradiance sub-bursts}
\label{sec:sim}
We simulate the intrinsic emission using the formalism described in Section~\ref{sec:super}, adopting the 21-cm atomic hydrogen transition, although the choice is not necessarily limited to it. Other molecular transitions capable of sustaining a population inversion, such as the 1612/1665/1667/1712 and 6030~MHz OH lines \citep{Houde2018, houde2019triggered}, are also candidates within this framework. We vary several physical parameters, such as the number density, relaxation and dephasing timescales, the strength and duration of the triggering pump signal, to generate a set of 15 distinct SR pulses. These pulses differ in duration, pulse shape, and delay time and are constructed to reproduce the characteristic timescales of a typical FRB emission. The resulting diversity of intrinsic profiles is reflected in Figure~\ref{fig:sr_profile}.

To generate sub-bursts from individual pulses, we apply the relativistic Doppler transformation defined in Equation~(\ref{eq:velspread}) to shift the intrinsic 21-cm hydrogen line ($\nu^\prime \approx 1420$~MHz) into different observing bands. In particular, we consider four central observing frequencies: 800~MHz, 1420~MHz, 4660~MHz, and 6030~MHz. The bandwidth of each band is determined using Equation~(\ref{eq:bw_1}). Observational measurements, however, exhibit some scatter around the nominal value of $B_\nu = 0.16$. To account for this variability, we introduce a Gaussian scatter with $\sigma = 0.01$ in our simulations to reproduce the statistical spread in the observed bandwidths. This procedure generates the wideband structure characteristic of a typical sub-burst, yielding a total of 60 sub-bursts, 15 in each of the four frequency bands. 

We reiterate that the TRDM is not restricted to a specific emission mechanism and may be applied to any model capable of producing a narrowband emission that exhibits the features of FRBs. Thus, any pulse emitted at some central frequency can be transformed to any desired frequency using Doppler shift, assuming relativistic motions in the source. Our SR-based emission model (Section ~\ref{sec:super}) incorporates a trigger and a delay and naturally fits the framework of TRDM. 

We subsequently apply five distinct scattering constants: $\Lambda_{\mathrm{sc, true}} = $ 2 ms, 5 ms, 10 ms, 15 ms, 20 ms (at 1 GHz) into each sub-bursts through Equation (\ref{eq:conv_1}). Together with the unscattered case, this results in a total of 360 simulated bursts. White Gaussian noise is then added to each burst to model random instrumental and thermal noise present in radio observations. A representative subset of 24 sub-bursts is shown in Figure~\ref{fig:sub_tile}. 

Up to this stage, our pipeline produces bursts with and without scattering while assuming a residual dispersion measure of $\Delta \mathrm{DM} = 0.0$ \pcm. To incorporate residual dispersion, we use the dedispersion routine in {\scshape FRBGUI} \citep{chamma2023broad} to inject a grid of $\Delta \mathrm{DM}$ values spanning $-5.0$~\pcm~to $5.0$~\pcm~in steps of $0.5$~\pcm. We then repeat the measurements of the sub-burst slope and duration for each sub-burst at every injected $\Delta \mathrm{DM}$ trial. 

Although certain parameters (e.g., the scattering timescale or residual dispersion) are known a priori for the simulated bursts, the methodology developed hereafter is entirely general and does not rely on prior knowledge of these quantities. The known parameters are used only a posteriori to validate and quantify the performance of our methods.

\subsection{Fits to individual sub-bursts}
\label{sec:fits}
\subsubsection{Burst pre-processing}
\label{sec:bpp}
To identify frequency channels that contain sufficient signal, we apply a Savitzky-Golay (SG) filter \citep{num_recip2007} as a pre-processing step. The filter suppresses high-frequency noise while preserving the underlying burst morphology, thereby augmenting the contrast between signal-bearing and noise-dominated channels and enabling more accurate channel identification. Here, the S/N is defined as the ratio of the peak amplitude to the standard deviation of the off-pulse noise. Channel selection is performed on the smoothed data using a fixed threshold of $\mathrm{S/N_{SG}}>10$ ensuring that only channels with sufficient signal are retained. 

Importantly, all measurements hereafter are carried out on the original (unsmoothed) data to avoid introducing biases in the inferred burst properties. The smoothing is used solely for detection and does not affect the quantitative analysis. Alternative channel-selection strategies may likewise be employed, provided they reliably distinguish intrinsic pulse emission from noise or interference.

\subsubsection{Sub-burst slope measurement}
\label{sec:slp_meas}

After pre-processing, the selected frequency channels are grouped into sets of four consecutive channels, and the intensities within each group are summed to produce a frequency-integrated pulse profile for each sub-band. We then fit the profile with our model, given in Equation (\ref{eq:con_fin}), using non-linear least-squares minimization implemented via the \textsc{\texttt{scipy}} package \citep{scipy}. Although these fits appear visually credible, non-linear multi-exponential models are intrinsically ill-conditioned due to parameter non-identifiability and degeneracy. In particular, when $\tau_{\mathrm{sc}} \approx t_{\mathrm{w}}$, the model parameters become strongly correlated, leading to non-unique solutions and unreliable uncertainty estimates. Our algorithm switches to fit Equation (\ref{eq:con_lim}) in this limit. However, our primary quantity of interest is the centroid of the pulse profile, a derived observable that remains well constrained by the data even when the underlying model parameters are not.

To get reliable estimates of centroid uncertainties, we marginalize over all model profiles consistent with the data using Goodman \& Weare’s Affine invariant Markov Chain Monte Carlo (MCMC) sampler implemented within the \textsc{\texttt{emcee}} package \citep{mcmc_2013}. For each sub-band, an ensemble of acceptable model realizations is generated, from which the posterior distribution of the centroid is computed. 

The centroids of the sub-bands are measured using
\begin{align}
     t_\mathrm{c}  = \frac{\int_0^\infty t \cdot I(\nu ,t) \ dt}{\int_0^\infty  I(\nu ,t) dt},
    \label{eq:t_c_gen}
\end{align}
following Paper~I. The mean of the centroid posterior is adopted as the centroid measurement, while the standard deviation provides the associated uncertainty. This approach naturally accounts for parameter degeneracies and yields reliable uncertainties on the derived quantity of interest. After obtaining centroids across all sub-bands spanning the burst bandwidth, we perform an orthogonal distance regression (ODR) \citep{scipy} to determine the sub-burst slope from the centroid trajectory. 

The burst duration, defined as the square root of the second central moment,
\begin{align}
     \lambda^2 (\nu) =  \frac{\int_0^\infty (t - t_\mathrm{c})^2 \cdot I(\nu ,t) \ dt}{\int_0^\infty  I(\nu ,t) \,dt},
    \label{eq:dur_gen}
\end{align}
is measured at the central frequency of the sub-burst as the mean of the corresponding MCMC posterior distribution. The uncertainty in duration is taken as the standard deviation of that distribution. These measurements are repeated across the $\Delta\mathrm{DM}$ values.

\subsection{Fits to the sub-burst slope law}
\label{sec:slp_law_fits}
After obtaining the measurements, we group the sub-bursts into six populations based on their injected scattering constant $\Lambda_{\mathrm{sc,true}}$. This classification rests on the astrophysical assumption that sub-bursts originating from a common progenitor traverse the same intervening medium, thereby experiencing a consistent scattering environment. The corresponding scattering timescale varies with frequency for a particular source, according to Equation (\ref{eq:tau_1}). 

Within each population, we model the (frequency-normalized) sub-burst slope using the modified law derived in Paper I,
\begin{align}
     \left\langle \frac{1}{\nu}\frac{d\nu}{dt_{\mathrm c}}\right\rangle_{ \nu_0}
   = -\frac{1}{\Delta\nu}
     \int_{\nu_0 - \Delta\nu/2}^{\nu_0 + \Delta\nu/2}
         \frac{\, d\nu}{ t_{\mathrm D} + t_{\mathrm w} + 2\Delta t_\mathrm{DM} + n\tau_\mathrm{sc}},
	\label{eq:slp_law_joint}
\end{align}
where we fix the scattering index to $n=4.0$ under the thin screen approximation. Throughout this work, ``sub-burst slope'' refers to the frequency normalized quantity, unless explicitly noted otherwise. 

The sub-burst duration is obtained by substituting Equation~(\ref{eq:con_fin}) into Equation~(\ref{eq:dur_gen}), yielding
\begin{align}
\lambda_\mathrm{c} = \sqrt{ t_{\mathrm{w}}^2(\nu_0) + \tau^2_{\mathrm{sc}}(\nu_0)}.
\label{eq:lam_joint}
\end{align}

For each population, we repeat the measurements over the aforementioned range of $\Delta \mathrm{DM}$. To ensure that the retained measurements remain physically consistent with the assumptions of the TRDM framework, we exclude sub-bursts that fail quality-control criteria. Specifically, we discard measurements for which no reliable fit is obtained, those with positive sub-burst slopes, those with relative slope uncertainties exceeding $40\%$, and those for which the measured normalized slope is not significantly larger than its associated uncertainty. These filtering criteria are implemented within FRBGUI \citep{chamma2023broad}, which is used to construct the measurement data frames used in our analysis. The remaining sub-burst slopes and durations are then jointly fit to Equations (\ref{eq:slp_law_joint}) and  (\ref{eq:lam_joint}) using the \texttt{scipy} non-linear least-squares package \citep{scipy}.

The simultaneous fitting of sub-burst slopes and durations across frequency helps break the degeneracy between $\tau_{\mathrm{sc}}$ and $t_{\mathrm{w}}$, leading to improved constraints on the scattering timescale, residual dispersion, intrinsic width, and other parameters entering Equation~(\ref{eq:slp_law_joint}). The recovered parameters are then compared with their true values to assess the accuracy of the method.

\section{Results and Discussion}
\label{sec:res}
Having constructed a population of simulated FRBs both with and without propagation effects, we now apply the modified sub-burst slope law to evaluate how accurately these quantities can be recovered from the observed burst properties. 

The fitting procedure treats the $A$-parameter (Equation~\ref{eq:a_value}), the intrinsic timescale $t_0 = \tau_{\mathrm{w}}' \nu'$ (as discussed after Equation~\ref{eq:velspread}), the residual dispersion ($\Delta \mathrm{DM}_{\mathrm{meas}}$), and the scattering constant ($\Lambda_{\mathrm{sc,meas}}$) as free parameters in Equation~(\ref{eq:slp_law_joint}). Hereafter, the true values are denoted by $\Delta \mathrm{DM}_{\mathrm{true}}$ and $\Lambda_{\mathrm{sc,true}}$, while the corresponding values recovered by our fitting procedure are denoted by $\Delta \mathrm{DM}_{\mathrm{meas}}$ and $\Lambda_{\mathrm{sc,meas}}$.  

The model parameters are initialized at $A_0 = 0.12$, $t_{0, 0} = 1500~\mathrm{ms\,MHz}$, $\Lambda_\mathrm{sc, 0} = 1.0~\mathrm{ms}$, and $\Delta \mathrm{DM}_0 = 0.0$. These parameters are constrained within physically motivated bounds: $0.05 \leq A \leq 0.30$, $1000 \leq t_0 \leq 2500~\mathrm{ms\,MHz}$, $0 \leq \Lambda_\mathrm{sc} \leq 50~\mathrm{ms}$, and $-5$~\pcm$ \leq \Delta \mathrm{DM} \leq 5$~\pcm. These limits ensure numerical stability and exclude unphysical solutions. We verify that the recovered parameters lie well within these bounds, indicating that the results are not driven by boundary effects.

For the remainder of the analysis, we set the bandwidth to $\Delta \nu _{\mathrm{obs}} \simeq 0.16\nu _{\mathrm{obs}}$ (Equation \ref{eq:bw_1}). We define $\nu _{\mathrm{obs}}$ as the center frequency of the selected frequency channels after burst-preprocessing. Tests performed by varying the proportionality constant $B_\nu$ in the range $0.1-0.2$ resulted in quantitatively similar parameter estimates and associated uncertainties.

\subsection{Interpretation of Fits in the Slope-Duration Plane}
We present our results in the sub-burst slope-duration plane, rather than as separate functions of frequency, as this representation provides a compact visualization of the burst evolution and enables direct comparison with previous studies. We note, however, that the fitted relations are not expected to pass exactly through the individual data points in this plane.

This behavior arises primarily because the fitting procedure is performed jointly in the sub-burst slope-frequency and duration-frequency domains, rather than directly in the slope-duration plane. In the standard TRDM framework, Equation~(\ref{eq:sub_burst}) implies a simple inverse dependence on the intrinsic timescale, with $t_{\mathrm{w}} \propto \nu^{-1}$, such that the sub-burst slope-duration relation directly characterizes the burst properties. However, in the presence of propagation effects, both the measured sub-burst slope and duration acquire additional frequency dependence through contributions from scattering and residual dispersion. The slope-duration relation therefore represents a projection of the underlying frequency-dependent fits, rather than the space in which the optimization is performed.

An additional source of scatter stems from the intrinsic diversity of the simulated burst population. Each population is constructed from 15 distinct SR profiles generated with different intrinsic parameters, producing variations in burst morphology and in quantities such as $A$ and $t_0$. Consequently, the fitted relations trace the aggregate behavior of the population rather than the properties of individual sub-bursts.

\begin{figure}
\centering
 \includegraphics[width=\columnwidth ]{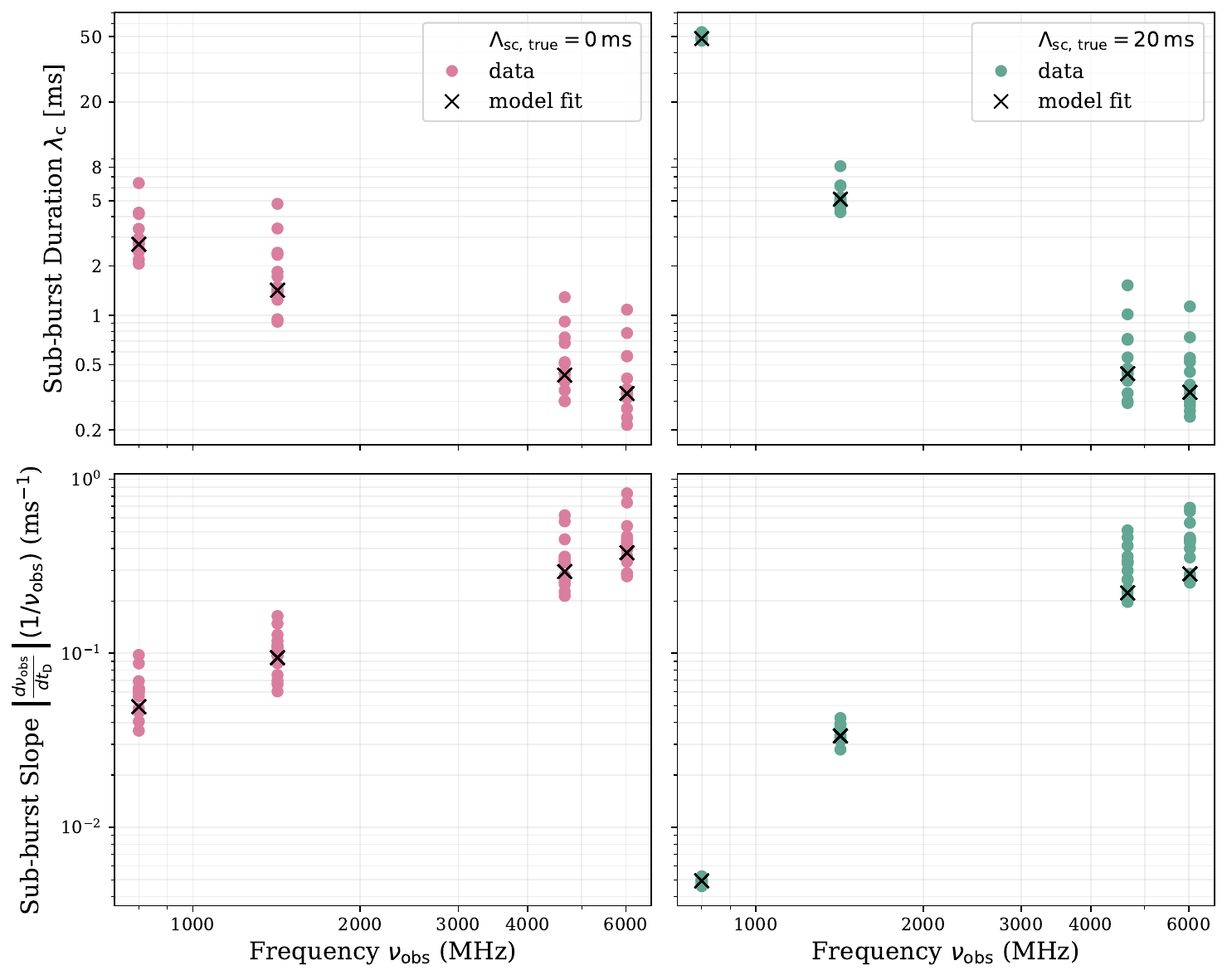}
 \caption{Sub-burst duration (top) and sub-burst slope (bottom) as a function of frequency for bursts without scattering (left panels) and a scattering constant $\Lambda_\mathrm{sc,true} = 20.0\;\mathrm{ms}$ (right panels). The simulated data are shown as solid circles, while the model fits to the data are indicated by black crosses. }
 \label{fig:datavmodel}
\end{figure}

To illustrate this, Figure~\ref{fig:datavmodel} shows the measured sub-burst durations and slopes as functions of frequency, i.e., in the space where the fitting procedure is performed. In the absence of scattering ($\Lambda_{\mathrm{sc,true}} = 0.0~\mathrm{ms}$), we observe a broad scatter of data points at a particular frequency due their intrinsically different SR profiles. When scattering is introduced ($\Lambda_{\mathrm{sc,true}} = 20.0~\mathrm{ms}$), it causes a strong frequency-dependent distortion of the profile that is most pronounced at lower frequencies, where it dominates over intrinsic timescales. This leads to a tighter clustering of sub-bursts and a closer alignment with the fitted relation. In contrast, at higher frequencies, where propagation effects are weaker and intrinsic properties are retained, the population exhibits a broader spread even for a large scattering timescale. The fitted relation (denoted by black crosses) therefore represents the aggregate behavior of the population, rather than the properties of individual sub-bursts. 

These considerations should therefore be kept in mind when interpreting the sub-burst slope-duration relations presented in Figures~\ref{fig:slp_law_joint}, \ref{fig:diff_dm}, and \ref{fig:scatt_dm}.


\subsection{Scattering-exclusive analysis}
\label{sec:scat_ex}
We begin by considering simulated sub-bursts that include scattering but no residual dispersion. Although dispersive delays are absent from the simulated data, $\Delta\mathrm{DM}_{\mathrm{meas}}$ is retained as a free parameter in the modified sub-burst slope law. This allows us to test whether the fitting procedure can reliably distinguish dispersive effects from other contributions, and to evaluate its likelihood of recovering non-zero $\Delta\mathrm{DM}_{\mathrm{meas}}$ in its absence.

\begin{figure*}[t]
\centering
 \includegraphics[width=0.8\textwidth ]{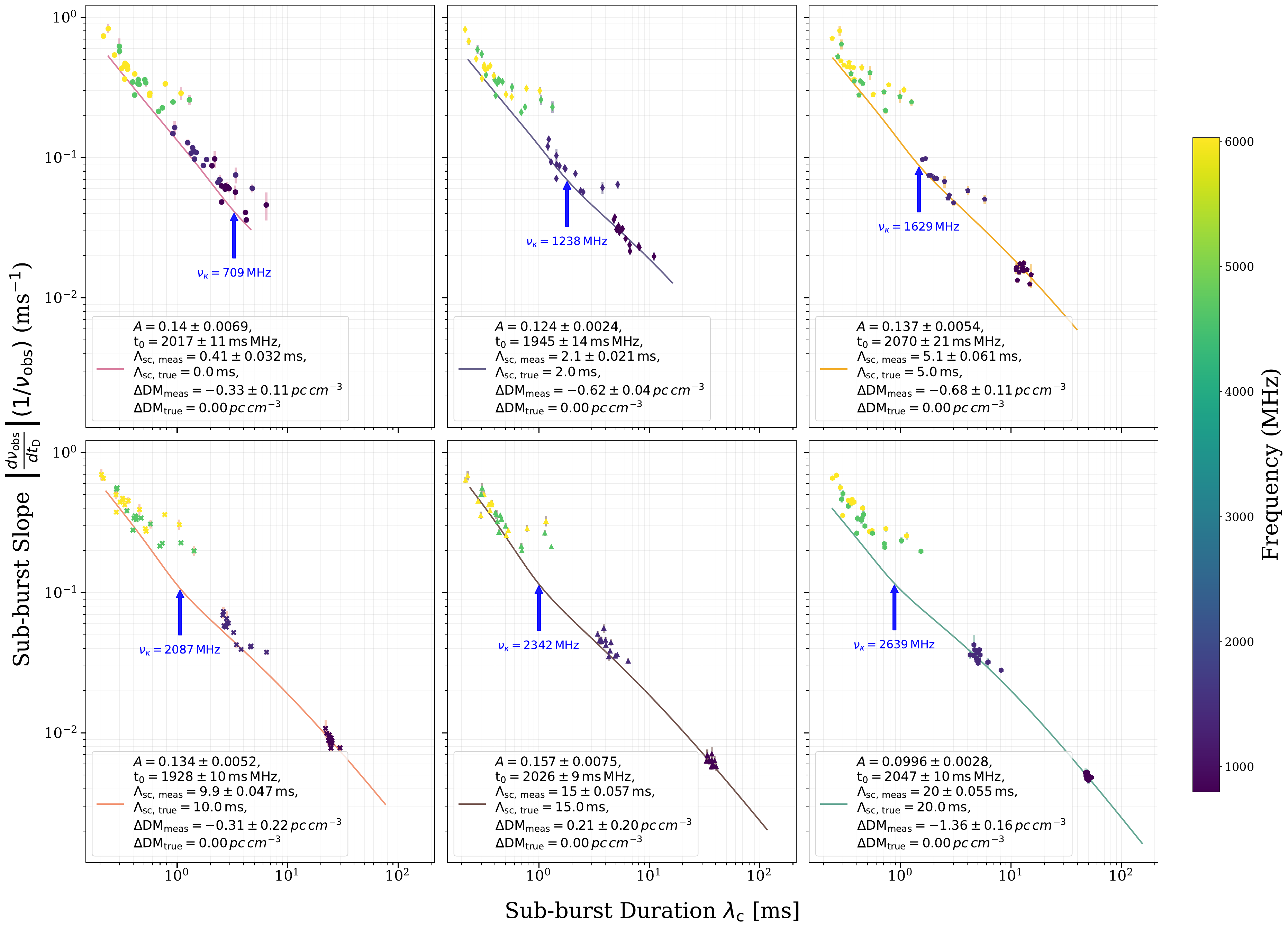}
 \caption{Fit to the modified sub-burst slope (Equation \ref{eq:slp_law_joint}) and duration (Equation \ref{eq:lam_joint}) following the procedure outlined in Section~\ref{sec:slp_law_fits}. The scattering timescale is varied, while no residual dispersion is present in the bursts. The legend reports the parameter values recovered from the fitting procedure. The true values of the residual dispersion and scattering timescale are denoted by the subscript “true,” while the measured values are labeled “meas.” The blue arrows reflect the frequency at which the slope-duration curve attains maximum curvature, as defined in Equation (\ref{eq:curvature}). }
 \label{fig:slp_law_joint}
\end{figure*}

Figure~\ref{fig:slp_law_joint} shows the modified sub-burst slope-duration relation for six populations constructed with different scattering constants. Each panel corresponds to a fixed value of $\Lambda_{\mathrm{sc,true}}$, with $\Delta \mathrm{DM_{true}}=0.0$~\pcm~in all cases. The most notable trend is the reduction in the normalized sub-burst slope with increasing $\Lambda_{\mathrm{sc,true}}$, which is more pronounced at low frequencies (or longer duration) where the scattering contribution is largest. In the scattering-dominated limit, characterized by $n\tau_{\mathrm{sc}} \gg t_{\mathrm D}+t_{\mathrm w}+2\Delta t_{\mathrm{DM}}$,
the slope evolution is governed primarily by scattering rather than by the intrinsic timescales. As $\Lambda_{\mathrm{sc,true}}$ increases, this condition is satisfied over an increasingly broad frequency range, producing a visible transition in the slope-duration relation from a scattering-subdominant to a scattering-dominated regime. This transition is indicated by the blue upward arrows in Figure~\ref{fig:slp_law_joint}. To identify its location, we define the transition frequency, $\nu_\kappa$, as the point at which the curvature of the trajectory in the logarithmic slope-duration plane is maximized. The curvature of a two-dimensional parametric curve is given by
\begin{align}
    \kappa = \frac{x'y'' - y'x''}{(x'^2 + y'^2)^{3/2}}
    \label{eq:curvature}
\end{align}
where $x = \log \lambda_{\mathrm{c}}$ and $y =\log \left[(1/\nu)\left|d\nu/dt_{\mathrm{c}}\right|\right]$ \citep{differential_geom}. Here, $x' \equiv dx/d\nu$, $y' \equiv dy/d\nu$, $x'' \equiv d^2x/d\nu^2$, and $y'' \equiv d^2y/d\nu^2$, with all derivatives evaluated numerically along the model curve. We find that $\nu_\kappa$ shifts to higher frequencies, or equivalently shorter durations, as $\Lambda_{\mathrm{sc,meas}}$ increases. For example, $\nu_\kappa \approx 1200~\mathrm{MHz}$ for $\Lambda_{\mathrm{sc,meas}}=2.1 \pm 0.021~\mathrm{ms}$, whereas $\nu_\kappa \approx 2600~\mathrm{MHz}$ for $\Lambda_{\mathrm{sc,meas}}=20.0 \pm 0.055~\mathrm{ms}$.

The scattering constant is recovered with high precision across all simulations, with a median absolute error of $0.07$~ms and a median absolute deviation (MAD) of $0.02$~ms, corresponding to a fractional accuracy of $1.3 \pm 0.9\%$. By contrast, the recovery of residual dispersion shows a systematic offset even though $\Delta \mathrm{DM_{true}} = 0$~\pcm. The median absolute error in $\Delta \mathrm{DM}_{\mathrm{meas}}$ is $0.48$~\pcm, with a MAD of $0.18$~\pcm, indicating that $\Delta \mathrm{DM}$ is only weakly constrained by the modified sub-burst slope law in the scattering-exclusive case.

The recovery of the $A$-parameter is more difficult to interpret because the simulated SR profiles are intrinsically diverse. Since no single intrinsic value of $A$ applies to the full population, the recovered value represents an effective population average with inherent scatter. Across the six simulated populations, the median recovered value of $A$ is $0.135$, with a MAD of $0.012$. For the largest scattering strength, $\Lambda_{\mathrm{sc,true}} = 20~\mathrm{ms}$, the recovered value of $A$ decreases to $0.0996$, a reduction of $\sim 26\%$ relative to the population median. This large variance in the recovered $A$ parameter, alongside the non-trivial recovery of a finite residual dispersion in its absence, can be understood in terms of the following effects:
\begin{enumerate}
    \item \textit{Relative importance of the scattering timescale:}
    As the scattering strength increases, the scattering-dominated regime extends to higher frequencies (as evident by increasing $\nu_\kappa$). A large fraction of sub-bursts therefore primarily constrain $\Lambda_\mathrm{sc}$, leaving fewer sub-bursts in the intrinsic regime where the slope law remains sensitive to $A$. The constraining power on $A$ is consequently reduced. In addition, scattering causes a reduction in the measured sub-burst slope magnitude across the band, including a mild suppression at the highest frequencies. The optimizer can partially accommodate this broadband reduction in slope magnitude by increasing the intrinsic drift timescale $t_\mathrm{D}$, or equivalently by decreasing $A$ (refer Equation \ref{eq:a_value}) in Equation (\ref{eq:slp_law_joint}). As a result, the recovered value of $A$ becomes systematically lower at large scattering timescales, despite the accurate recovery of $\Lambda_\mathrm{sc}$.

    \item \textit{Degeneracy between $\Delta \mathrm{DM}$ and the $A-$parameter:}
    The $\nu^{-1}$ and $\nu^{-2}$ contributions of $t_\mathrm{D}$ and $\Delta t_\mathrm{DM}$, which constrain $A$ and $\Delta \mathrm{DM}$, respectively, can partially mimic one another in the slope-law denominator (Equation~\ref{eq:slp_law_joint}) over a finite bandwidth, resulting in an inherent degeneracy between them. In the absence of true dispersive delays, $\Delta \mathrm{DM}$ becomes a redundant parameter, leading to an over-parameterized model in which the fit explores coupled $A$-$\Delta \mathrm{DM}$ solutions that produce comparable residuals, rather than converging to $\Delta \mathrm{DM} = 0$~\pcm. In particular, relatively steep slopes in the mid-to-high frequency regime (2.5-6 GHz) may be reproduced by an over-dedispersed solution ($\Delta \mathrm{DM} < 0$), while a reduced value of $A$ partially compensates for this steepening by increasing $t_\mathrm{D}$ (Equation \ref{eq:a_value}). This combination can yield a better global fit across the band than a zero-DM solution.  If we remove the $\Delta t_\mathrm{DM}$ term in Equation~(\ref{eq:slp_law_joint}), an equivalent steepening would instead be achieved by increasing $A$, which reduces the intrinsic delay $t_\mathrm{D}$ and thus the overall denominator. However, because scattering still produces a global reduction in the slope magnitude, the recovered $A$ does not return fully to the intrinsic population median. Thus, in the reduced model, $A$ remains slightly biased at large scattering timescales, while in the full model the additional freedom in $\Delta \mathrm{DM}$ makes the two parameters non-unique.
\end{enumerate}

\subsubsection{TRDM sub-burst slope law under scattering}
\label{sec:A_value}
In Paper~I, we showed that the proportionality constant in the standard sub-burst slope law (Equation~\ref{eq:sub_burst}) is not invariant when the propagation effects alter the burst morphology. Slope measurements that depend on the full observed profile, including the centroid-trajectory slopes used here, can therefore be biased by propagation-induced broadening or distortion. To illustrate this behavior, we evaluate the relation for bursts affected by scattering, but without residual dispersion, and examine the resulting frequency dependence of an effective proportionality constant. We denote this constant by $A^{\ast}$ to distinguish it from the intrinsic parameter $A$ of the emission model. The effective, frequency-averaged value is defined as $A^{\ast}_\mathrm{eff}=\lambda_\mathrm{c}|s|$ (from Equation~\ref{eq:sub_burst}), where $\lambda_\mathrm{c}$ is the measured duration and $s$ is the sub-burst slope measured relative to the centroid (Equation \ref{eq:slp_law_joint}).

\begin{figure}
\centering
 \includegraphics[width=\columnwidth ]{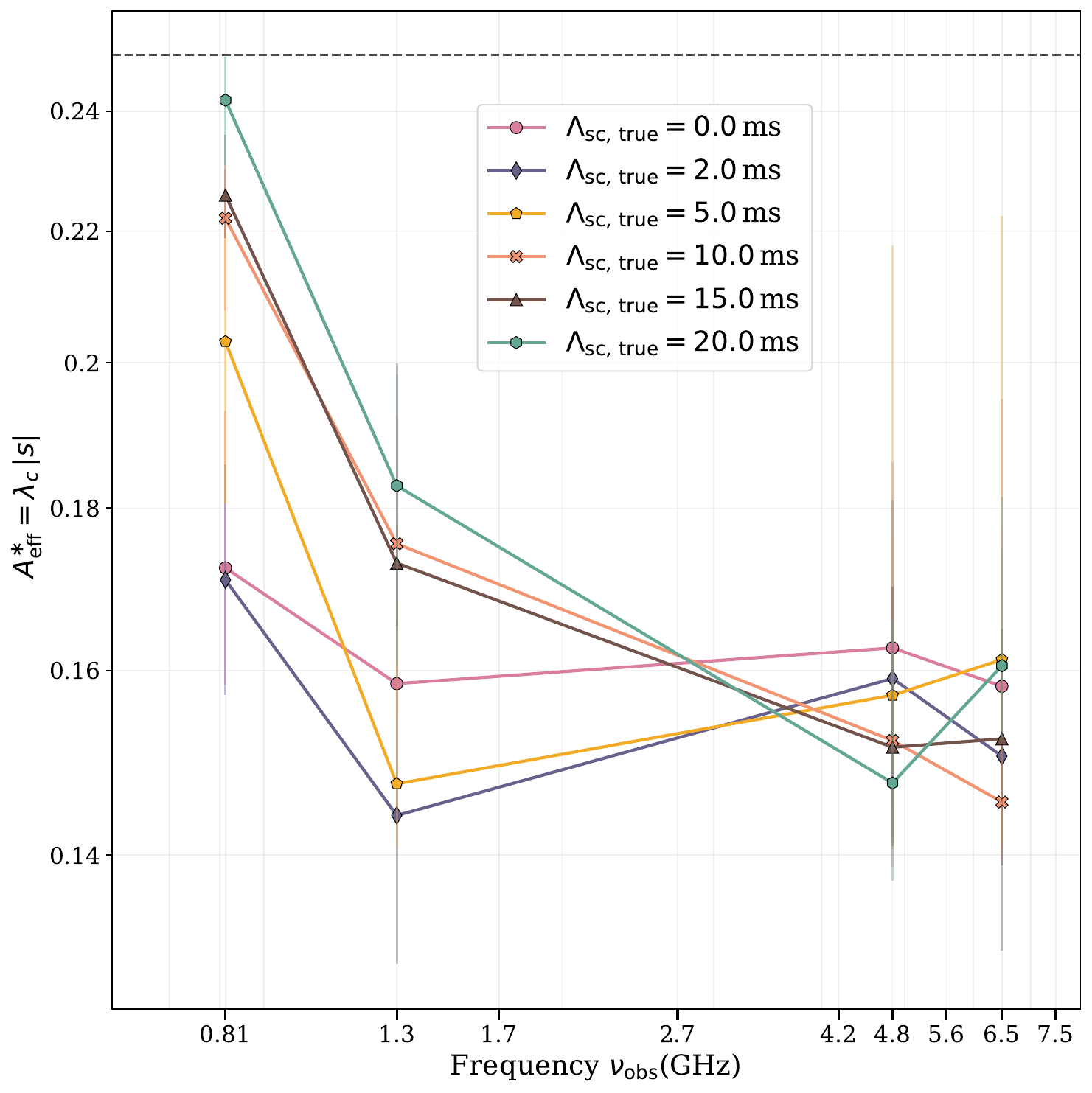}
 \caption{The median of the binned $A^\ast$-values for a frequency band against the mean frequency of that band for each scattering population. The dashed line is the constant $1/n\approx0.25$ for our analysis. The residual DM is zero for each population.}
 \label{fig:a_eff}
\end{figure}

Figure~\ref{fig:a_eff} shows the frequency-binned values of $A^{\ast}_\mathrm{eff}$, for different scattering timescales. A clear transition is observed between regimes in which scattering is sub-dominant and dominant. At high frequencies, where scattering is weak, $A^{\ast}_\mathrm{eff}$ approaches the intrinsic value of $A$. This behavior is consistent with the limiting form of Equation~(\ref{eq:slp_law_joint}), $s \propto A^{\ast}_\mathrm{eff}/t_{\mathrm w}$, valid for $A \ll 1$ and $t_{\mathrm w} \gg \tau_{\mathrm{sc}}$. In contrast, at low frequencies, $A^{\ast}_\mathrm{eff}$ asymptotically approaches $1/n \approx 0.25$, where $n$ is the scattering index defined in Equation~(\ref{eq:tau_1}). In this regime, Equation~(\ref{eq:slp_law_joint}) reduces to the scattering-dominated form, $s \propto 1/(n\tau_{\mathrm{sc}})$, for $\tau_{\mathrm{sc}} \gg t_{\mathrm w}$. Thus, these asymptotic behaviors arise naturally within the centroid-based formulation of the modified sub-burst slope law, whose limiting forms provide a direct means of quantifying the transition between intrinsically dominated behavior and behavior dominated by propagation effects. Alternative slope estimators may lead to different numerical values of $A^{\ast}_\mathrm{eff}$, reflecting differences in how the sub-burst slope law is defined and quantified. The centroid-based definition adopted here is advantageous as it captures this evolution through a simple and physically interpretable relation.

This result demonstrates that the proportionality constant inferred from the sub-burst slope law is inherently regime-dependent. Therefore, fitting a single TRDM relation across the full frequency range, without accounting for propagation effects, can lead to biased estimates of $A$. Furthermore, the observed frequency dependence of $A^{\ast}_\mathrm{eff}$ provides a direct diagnostic of the transition between intrinsic and propagation-dominated regimes, and highlights the necessity of incorporating scattering when interpreting the sub-burst slope law.

\subsection{Dispersion-exclusive analysis}

\citet{chamma2021evidence} highlighted that inaccuracies in the DM modify the apparent morphology of sub-bursts in the time-frequency plane, thereby altering the measured sub-burst slope while leaving the functional form of the inverse slope-duration relation (Equation \ref{eq:sub_burst}) largely intact. As a result, the proportionality constant $A$, which is inferred from these measurements, can also be biased by residual DM errors, causing the recovered value to deviate from its intrinsic source-frame value. We seek to understand this behavior and ascertain whether our algorithm is able to reliably disentangle dispersive effects from the intrinsic burst dynamics. We begin by considering the dispersion-only limit of the modified slope law by removing the scattering term from Equation~ (\ref{eq:slp_law_joint}). This avoids over parameterization as seen in  Section \ref{sec:scat_ex} and allows the residual dispersion contribution to be examined independently. The resulting expression (originally derived in Paper~I, Section~3.3) is
\begin{align}
     \left\langle \frac{1}{\nu}\frac{d\nu}{dt_{\mathrm c}}\right\rangle_{ \nu_0}
   = -\frac{1}{\Delta\nu}
     \int_{\nu_0 - \Delta\nu/2}^{\nu_0 + \Delta\nu/2}
         \frac{\, d\nu}{ t_{\mathrm D} + t_{\mathrm w} + 2\Delta t_\mathrm{DM} },
	\label{eq:slp_law_dm}
\end{align}
The sub-burst duration in the absence of scattering is simply
\begin{align}
\lambda_\mathrm{c} = t_{\mathrm{w}}(\nu_0) .
\label{eq:lam_dm}
\end{align}

\begin{figure*}[t]
\centering
 \includegraphics[width=0.8\textwidth ]{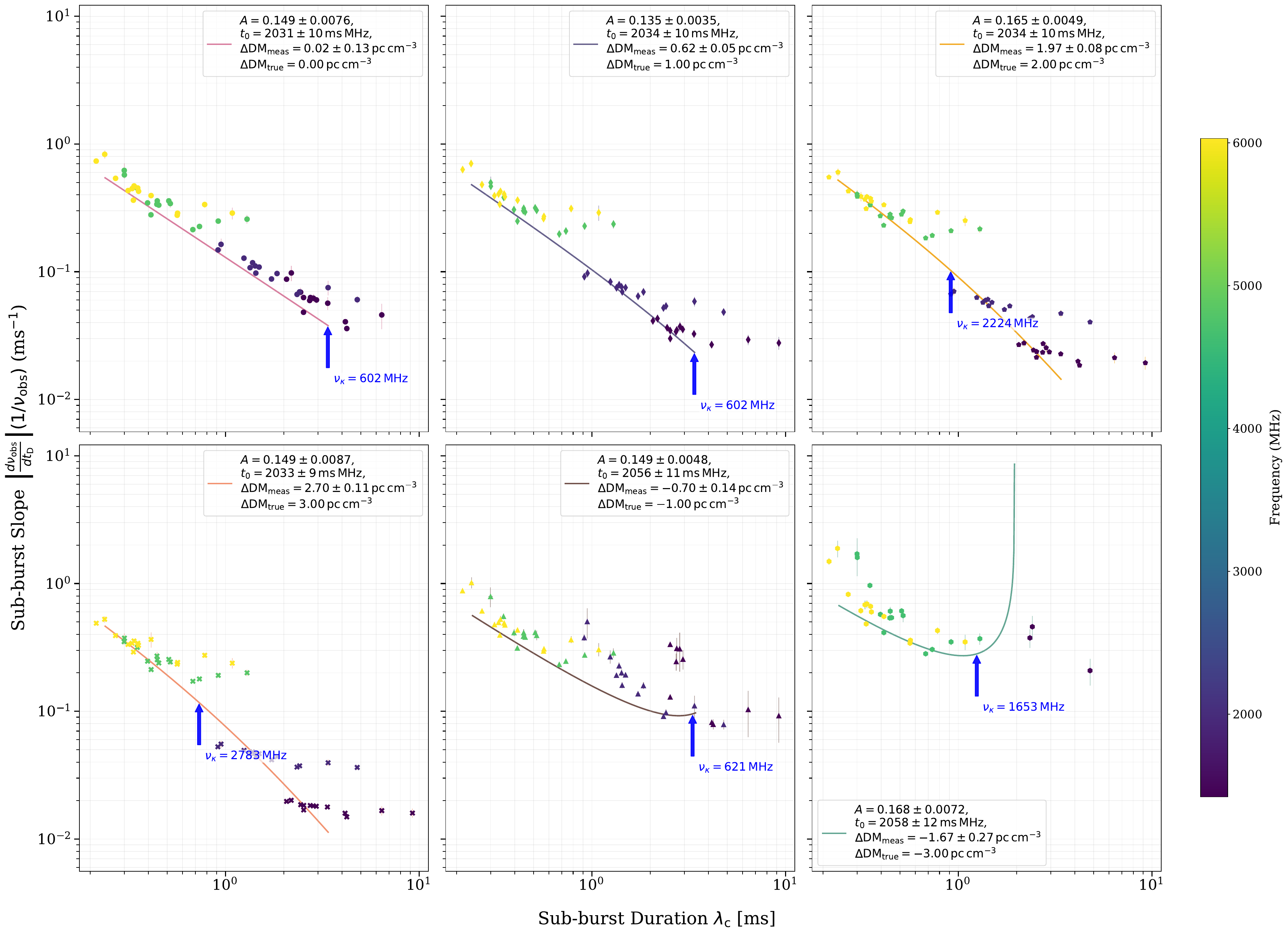}
 \caption{Fits to the modified sub-burst slope (Equation \ref{eq:slp_law_dm}) and duration (Equation \ref{eq:lam_dm}) for simulated bursts at varying residual dispersion measures $\Delta\mathrm{DM_{true}}$. The legend reports the true and recovered residual dispersion measures, $\Delta\mathrm{DM_{true}}$ and $\Delta\mathrm{DM_{meas}}$, along with the recovered sub-burst slope law parameter $A$. The frequency at which the slope-law attains maximum curvature is denoted by the blue arrow. }
 \label{fig:diff_dm}
\end{figure*}

Figure~\ref{fig:diff_dm} presents fits to the modified slope law and duration relations described by Equations~(\ref{eq:slp_law_dm}) and (\ref{eq:lam_dm}) for simulated bursts spanning $-3.0$~\pcm $\leq \Delta \mathrm{DM_{true}} \leq$ $+3.0$~\pcm. Although the full simulation grid extends to $\pm 5.0$~\pcm, strongly over-dedispersed cases with $\Delta \mathrm{DM_{true}} < -3.0$~\pcm~ frequently yield positive slopes, causing a significant fraction of the simulated bursts to be rejected by the measurement pipeline. The blue arrows mark the characteristic frequency, $\nu_\kappa$, defined using Equation~(\ref{eq:curvature}) as the point of maximum curvature along the plotted modified slope law in the slope-duration plane. For both types of dispersion, $\nu_\kappa$ shifts toward higher frequencies as $|\Delta \mathrm{DM}|$ increases.

The fitting procedure is sensitive to both the $\nu^{-2}$ dependence and the sign of the dispersive delay. Across the range of $\Delta \mathrm{DM_{true}}$ values, the residual dispersion is recovered with a median absolute error of $0.30$~\pcm~and a MAD of $0.18$~\pcm. We observe that under-dedispersion ($\Delta \mathrm{DM_{true}}>0$~\pcm) reduces the magnitude of the negative slope, causing the corresponding slope-law curves to bend downward relative to the $\Delta \mathrm{DM_{true}}=0$~\pcm~ relation. This downward offset becomes more pronounced for larger positive values of $\Delta \mathrm{DM_{true}}$ and toward lower observing frequencies, where the dispersive delay is strongest. On the other hand, over-dedispersion ($\Delta \mathrm{DM_{true}}<0$) steepens the slopes causing an upward curvature of the slope law, and can eventually drive the slopes to positive, unphysical values. Here, as $\Delta \mathrm{DM_{true}}$ becomes more negative, the frequency at which the slope law develops an upward curvature shifts toward higher frequencies. 

We find that the recovery of $\Delta \mathrm{DM}$ depends on the sign and magnitude of the residual dispersive offset. This asymmetry is most evident at the largest offsets. For example, the $\Delta \mathrm{DM_{true}}=3.0$~\pcm~ case is recovered as $\Delta \mathrm{DM_{meas}}=2.7\pm 0.11$~\pcm, corresponding to an absolute discrepancy of $\sim 0.30$~\pcm. By contrast, the $\Delta \mathrm{DM_{true}}=-3.0$~\pcm~ case is recovered as $\Delta \mathrm{DM_{meas}}=-1.67\pm 0.27$~\pcm~ yielding a substantially larger discrepancy of $\sim 1.33$~\pcm. The parameter $t_0$ remains remarkably stable across all trials, while the inferred value of $A$ exhibits moderate variations with $\Delta \mathrm{DM}$. The $A$ value is slightly elevated when $\Delta \mathrm{DM_{true}} = -3.0$~\pcm.

The weaker recovery in the over-dedispersed regime arises from two related effects. First, over-dedispersion introduces a delay term with the opposite sign to the intrinsic timescales, artificially steepening the sub-burst slope. When $2|\Delta t_\mathrm{DM}| \gtrsim t_\mathrm{D} + t_\mathrm{w}$, the denominator of Equation~(\ref{eq:slp_law_dm}) can change sign, causing the measured sub-burst slope to become positive. Since these points are excluded from the analysis, the fit is performed on a truncated subset of the over-dedispersed population, removing the low-frequency measurements where the $\nu^{-2}$ dispersion signature is strongest. Second, among the retained measurements, the few lowest-frequency surviving points, particularly those near $1420~\mathrm{MHz}$, provide the strongest remaining leverage on $\Delta \mathrm{DM_{meas}}$. These bursts retain a sufficiently strong intrinsic contribution to prevent the slope-law denominator from changing sign, such that $t_\mathrm{D} + t_\mathrm{w} > 2 |\Delta t_\mathrm{DM}|$. Their slopes therefore do not cleanly trace the $\nu^{-2}$ dependence of the imposed residual dispersion, but instead reflect the combined influence of the intrinsic and dispersive terms in the denominator of Equation~(\ref{eq:slp_law_dm}). Since the fit constrains this combined denominator rather than the two contributions independently, different combinations of $A$ and $\Delta \mathrm{DM}$ can reproduce similar slope-duration measurements. The optimizer therefore selects a less negative effective $\Delta \mathrm{DM_{true}}$ and a larger $A$, which together best reproduce the steepening seen in the retained population.

In the case of under-dedispersed bursts, the residual dispersive delay has the same sign as the intrinsic timescales, $t_{\mathrm D}$ and $t_{\mathrm w}$, and therefore contributes additively to the denominator of Equation~(\ref{eq:slp_law_dm}). This produces a smooth low-frequency flattening of the slope relation while keeping the denominator positive and away from zero, such that $t_\mathrm{D} + t_\mathrm{w} + 2 \Delta t_\mathrm{DM} >0$. The affected measurements are therefore retained in the fit and are dispersion dominated, providing the low-frequency leverage needed to identify the $\nu^{-2}$ contribution. Under-dedispersed cases are consequently recovered more accurately.

Since scattering is absent from both the simulated sub-bursts and the slope law, the model contains no additional broadening term that could act as an extra degree of freedom and absorb changes in the sub-burst slope and duration. Removing the scattering contribution therefore avoids over-parameterization and allows the dispersion-induced modifications to the slope law to be isolated more cleanly. This analysis demonstrates the utility of fitting reduced forms of the full modified slope law (Equations \ref{eq:slp_law_joint} and \ref{eq:lam_joint}). The flexibility of our framework allows scattering and dispersion effects to be included or excluded straightforwardly. For a detailed derivation of these reduced forms, see Paper~I. This allows us to assess whether a simplified model is sufficient or whether the full scattering-dispersion formulation, discussed in the following section, is required to obtain reliable parameter constraints.

\subsection{Joint scattering-dispersion analysis}
\begin{figure*}
\centering
 \includegraphics[width=\textwidth ]{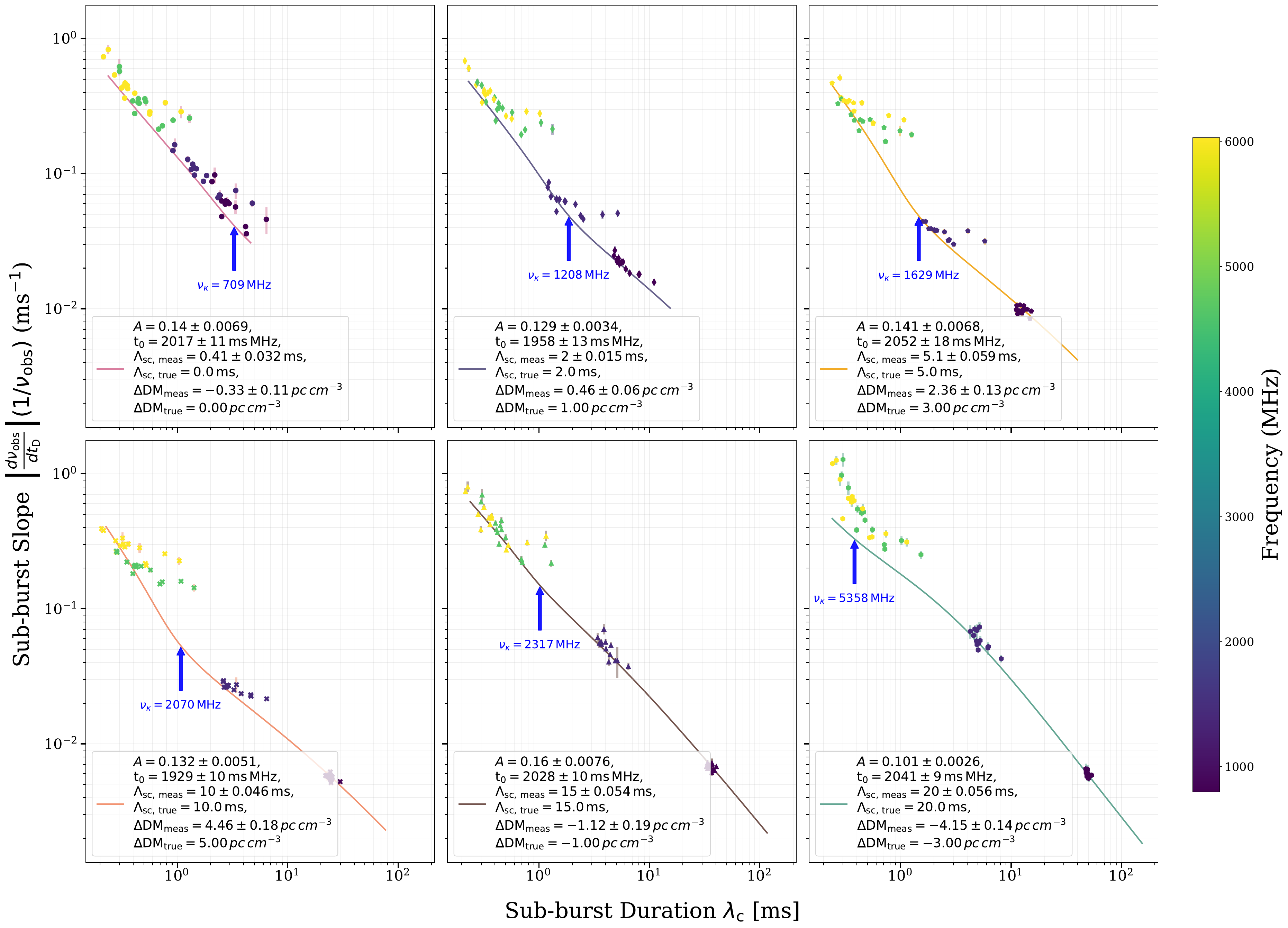}
 \caption{Same as Figure \ref{fig:slp_law_joint} but with varying scattering timescales and residual dispersion measures.  }
 \label{fig:scatt_dm}
\end{figure*}

Finally, in Figure~\ref{fig:scatt_dm}, we evaluate the performance of the modified sub-burst slope law, given by Equation~(\ref{eq:slp_law_joint}), for bursts containing signatures of both scattering and residual dispersion. We consider residual dispersion values spanning $-3.0$~\pcm $\leq \Delta \mathrm{DM_{true}} \leq$ $+5.0$~\pcm~ and scattering timescales in the range $0 \leq \Lambda_{\mathrm{sc,true}} \leq 20~\mathrm{ms}$. The blue arrows mark $\nu_\kappa$, the maximum-curvature point of the plotted slope-duration relation. These arrows therefore identify the transition from the approximately intrinsic inverse relation between slope and duration to the nonlinear behavior introduced by propagation effects. Relative to sub-bursts with no imposed scattering or residual dispersion, shown in the upper-left panel, these transitions shift toward higher frequencies as $\Lambda_\mathrm{sc,true}$ and $|\Delta \mathrm{DM_{true}}|$ increases. Additionally, as $\Lambda_{\mathrm{sc,true}}$ increases, the low-frequency sub-bursts cluster at longer durations, with $\lambda_{\mathrm c}$ approaching $\tau_{\mathrm{sc}}$ in the scattering-dominated limit, $\tau_{\mathrm{sc}} \gg t_{\mathrm w}$.

The scattering timescale is recovered with high precision, with a median absolute error of $0.088~\mathrm{ms}$ and a MAD of $0.041~\mathrm{ms}$. The residual dispersion measure is recovered with a median absolute error of $0.54$~\pcm~ and a MAD of $0.154$~\pcm. While the algorithm exhibits stable recovery of $A$ and $\Delta \mathrm{DM_{true}}$ in the under-dedispersed bursts, inconsistencies emerge in the over-dedispersed regime.

This apparent contrast between the two dispersion regimes can be understood in the context of preceding discussions. In the case of under-dedispersion ($0$~\pcm~$ < \Delta \mathrm{DM} \leq$$~5.0$~\pcm), the $\nu^{-2}$ dispersive delay contributes additively in the denominator of the modified slope law (Equation \ref{eq:slp_law_joint}), in the same sense as the scattering term. It therefore shifts the relation downward toward flatter slopes relative to the case with no propagation effects, while avoiding the asymptotic behavior encountered in the over-dedispersed regime (Figure \ref{fig:diff_dm}). The fitting procedure remains stable for these populations, recovering a mean value of $A = 0.134 \pm 0.006$. Although degeneracy between $A$ and $\Delta \mathrm{DM}$ produces a modest loss of sensitivity in their estimation, it does not strongly bias the recovered parameters. The residual dispersion is constrained to within $\sim 0.5$~\pcm~ of the true value, and the corresponding adjustments in the slope law are absorbed by small shifts in $A$, which remains confined to a narrow range, $A\sim0.13$--$0.14$. This loss is subtly enhanced by scattering, which reduces the slope at all frequencies.  

For moderate over-dedispersion, the algorithm retains partial sensitivity to the intrinsic behavior through the high-frequency sub-bursts, where propagation effects are weaker. However, when strong over-dedispersion is combined with a large scattering timescale, the intrinsic parameters become much harder to constrain. In the $\Delta \mathrm{DM} = -3.0~$\pcm, $\Lambda_\mathrm{sc} = 20$~ms case (shown in the bottom-right panel of Figure~\ref{fig:scatt_dm}), the recovered value of $A$ is significantly lower ($A = 0.101 \pm 0.0026$), about $26\%$ below the median value of $A=0.136$ recovered across the six populations. The fitted residual DM is also biased toward a more negative value, $\Delta \mathrm{DM_{meas}} = -4.15 \pm 0.14$~\pcm, compared with the true value of $\Delta \mathrm{DM_{true}} = -3.0$~\pcm. 

The breakdown of parameter recovery in the over-dedispersed, high-scattering regime, can be attributed to the lack of undistorted bursts that directly constrain the intrinsic parameter $A$. For the plotted best-fit modified slope law, $\nu_\kappa$ shifts to $5358~\mathrm{MHz}$, indicating that the inferred slope-duration relation is strongly dominated by propagation effects. Although this transition frequency may differ slightly from that obtained by fixing $\Delta \mathrm{DM}$ to its true value, it shows that only the highest-frequency sub-bursts remain in the region of the slope-duration relation where the intrinsic timescales contribute appreciably. Consequently, most of the observed slope-duration relation is governed by scattering and residual dispersion, leaving the global fit only weakly sensitive to changes in $A$. This reduced leverage on $A$ is further compounded by its degeneracy with $\Delta \mathrm{DM}$. As discussed in the preceding subsections, different combinations of these parameters can reproduce similar slope-duration measurements over the available frequency range. For this regime, the fit favors a more negative value of $\Delta \mathrm{DM}$, which steepens the predicted slope, together with a reduced value of $A$, which partially compensates for this effect by increasing the intrinsic delay term. The recovered parameters therefore represent the best global compromise permitted by the coupled $A -\Delta \mathrm{DM}$ degeneracy in a population dominated by propagation effects. Reliable recovery of intrinsic parameters in over-dedispersed samples therefore requires sufficient measurements at frequencies where the intrinsic timescales dominate over both the scattering and residual dispersive terms.

\section{Summary}
\label{sec:sum}

We have developed and validated an analytical framework to disentangle scattering and dispersion effects from the intrinsic emission properties of repeating FRB sub-bursts. We generate simulated bursts within the TRDM framework, assuming superradiance as the underlying emission mechanism, and incorporate a broad range of scattering timescales and residual dispersion measures. For each sub-burst, we measure the slope from the centroid evolution in the dynamic spectrum and the characteristic duration at the center frequency of the sub-burst profile using parametric fitting. Then, by fitting the modified sub-burst slope law to the measured slope-duration pairs, we simultaneously recover intrinsic emission parameters, scattering timescales, and residual dispersion measures. 

We divide our analyses across three cases. In the scattering-exclusive analysis, the scattering timescale is recovered with appreciable accuracy. However, as $\Delta \mathrm{DM}$ is retained as a free parameter even when no residual dispersion is present in the sub-bursts, it results in an over-parameterized model. This, together with the degeneracy between $A$ and $\Delta \mathrm{DM}$, leads to comparatively weak constraints on both parameters.

In the dispersion-exclusive case, removing the scattering contribution alleviates over-parameterization and reveals an asymmetry in parameter recovery between the under- and over-dedispersed bursts. Under-dedispersed bursts are recovered more reliably as the residual dispersive delay adds to the intrinsic timescales, creating a continuous slope-duration relation. Over-dedispersion instead subtracts from the intrinsic terms, and can cause the denominator of the sub-burst slope to be zero or even positive. These positive slopes are unphysical and are excluded from our model. This reduction in usable measurements, combined with the $A-\Delta \mathrm{DM}$ degeneracy, weakens the recovery of both parameters in this regime.

When both propagation effects are varied simultaneously in the joint scattering-dispersion case, parameter recovery remains sensitive to the type of residual dispersion, and is further influenced by scattering. Although the scattering timescale is precisely recovered across all regimes, parameters in the under-dedispersed regime are consistently better constrained than those in the over-dedispersed regime. This discrepancy peaks in the strongly scattered, strongly over-dedispersed regime, where the bulk of the emission is dominated by propagation effects, leaving only a small subset of high-frequency sub-bursts partially sensitive to $A$. Consequently, the recovered parameters reflect a global compromise dictated by the inherent $A\text{-}\Delta \mathrm{DM}$ degeneracy, rather than a clean recovery of the imposed values.  Extending the data set to even higher frequencies is therefore required to gain adequate leverage on $A$.

These results demonstrate that our framework provides a powerful tool to detect and disentangle foreground propagation effects from the intrinsic properties of FRB sources under controlled conditions. The flexibility of our model allows nuisance parameters to be easily omitted, thereby improving constraints on the primary parameters of interest and avoiding overparameterization. This adaptability is particularly advantageous when confronting real observations, which exhibit complexities such as  varying intrinsic properties, limited sample sizes, finite spectro-temporal resolution, variable $\mathrm{S/N}$ ratios, methodological and temporal $\mathrm{DM}$ variations, and limited frequency coverage. In Paper~III, we will extend this analysis to observational datasets to assess how these factors impact parameter recovery and the overall performance of our methods.

\section*{CRediT authorship contribution statement}
\textbf{Aishwarya Kumar:} Conceptualization, Methodology, Software, Formal analysis, Writing - Original Draft. \textbf{Fereshteh Rajabi:} Writing - Review \& Editing, Supervision, Funding acquisition. \textbf{Martin Houde:} Conceptualization, Writing - Review \& Editing, Supervision, Funding acquisition.

\section*{Acknowledgments}
M.H.’s research is funded through the Natural Sciences and Engineering Research Council of Canada (NSERC) Discovery Grant RGPIN-2024-05242.  F.R.'s research is supported by the NSERC Discovery Grant RGPIN-2024-06346.

\section*{Declaration of generative AI and AI-assisted technologies in the manuscript preparation process}
During the preparation of this work, the authors used ChatGPT to improve readability through grammar and language refinement and to assist with Python code for positioning plot annotations and inset elements. The authors reviewed and edited the output as needed and take full responsibility for the content of the publication.

\appendix

\section{Spectro-temporal diversity of the simulated FRB population}

\begin{figure*}
\centering
 \includegraphics[width=\textwidth ]{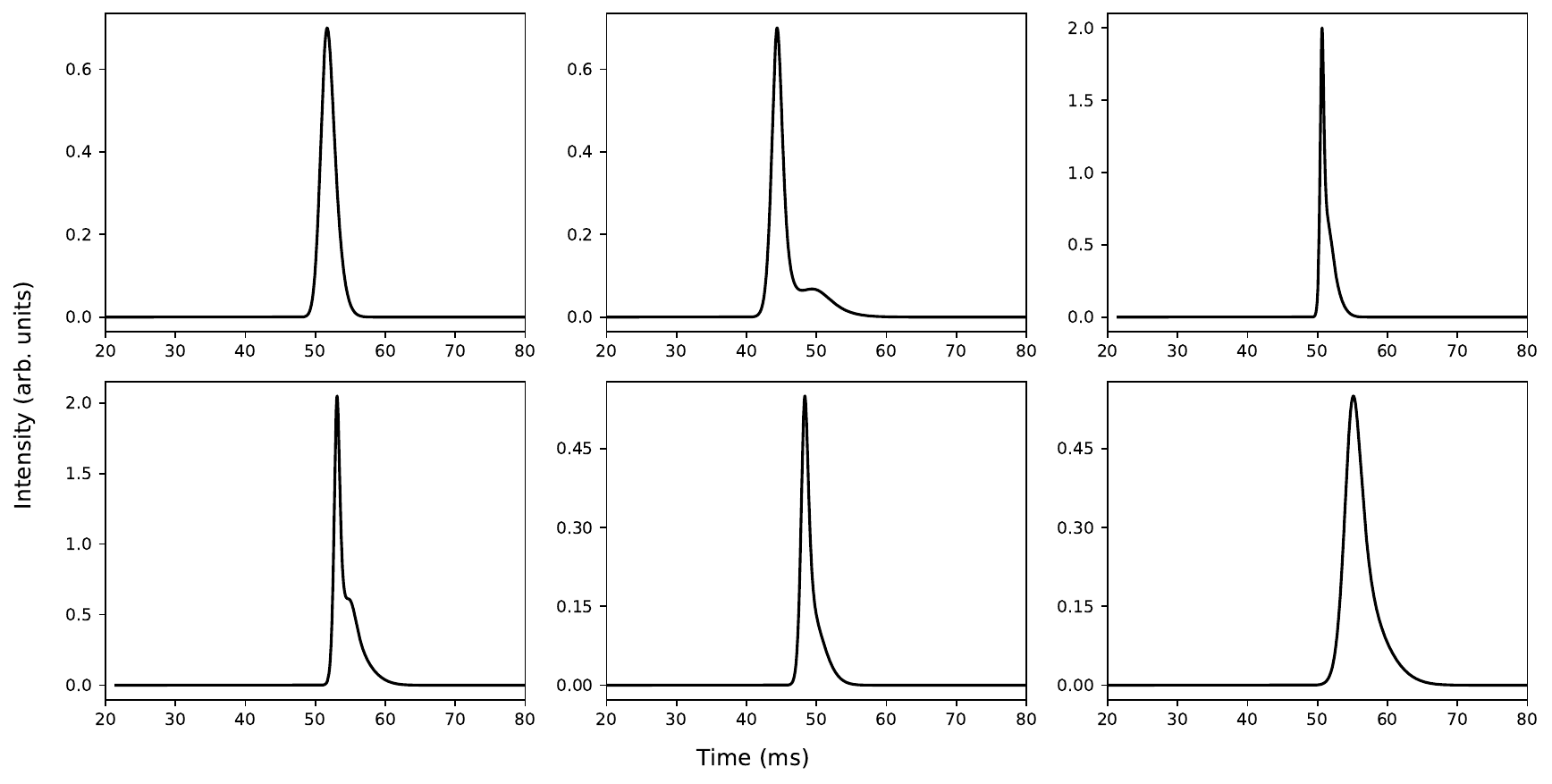}
 \caption{ A subset of the superradiant emission profiles used to generate the simulated FRB population at 1.4~GHz, illustrating the diversity in burst morphology arising from differences in the intrinsic emission parameters. }
 \label{fig:sr_profile}
\end{figure*}

In Figure~\ref{fig:sr_profile}, we illustrate the diversity of intrinsic burst profiles generated using the superradiance model. Each profile is produced with slightly different initial conditions, leading to variations in the intrinsic emission parameters. Consequently, the profiles exhibit a range of widths and morphologies, representative of the pulse-to-pulse variability observed in repeating FRB sources.

\label{sec:App}

\begin{figure*}
\centering
 \includegraphics[width=\textwidth ]{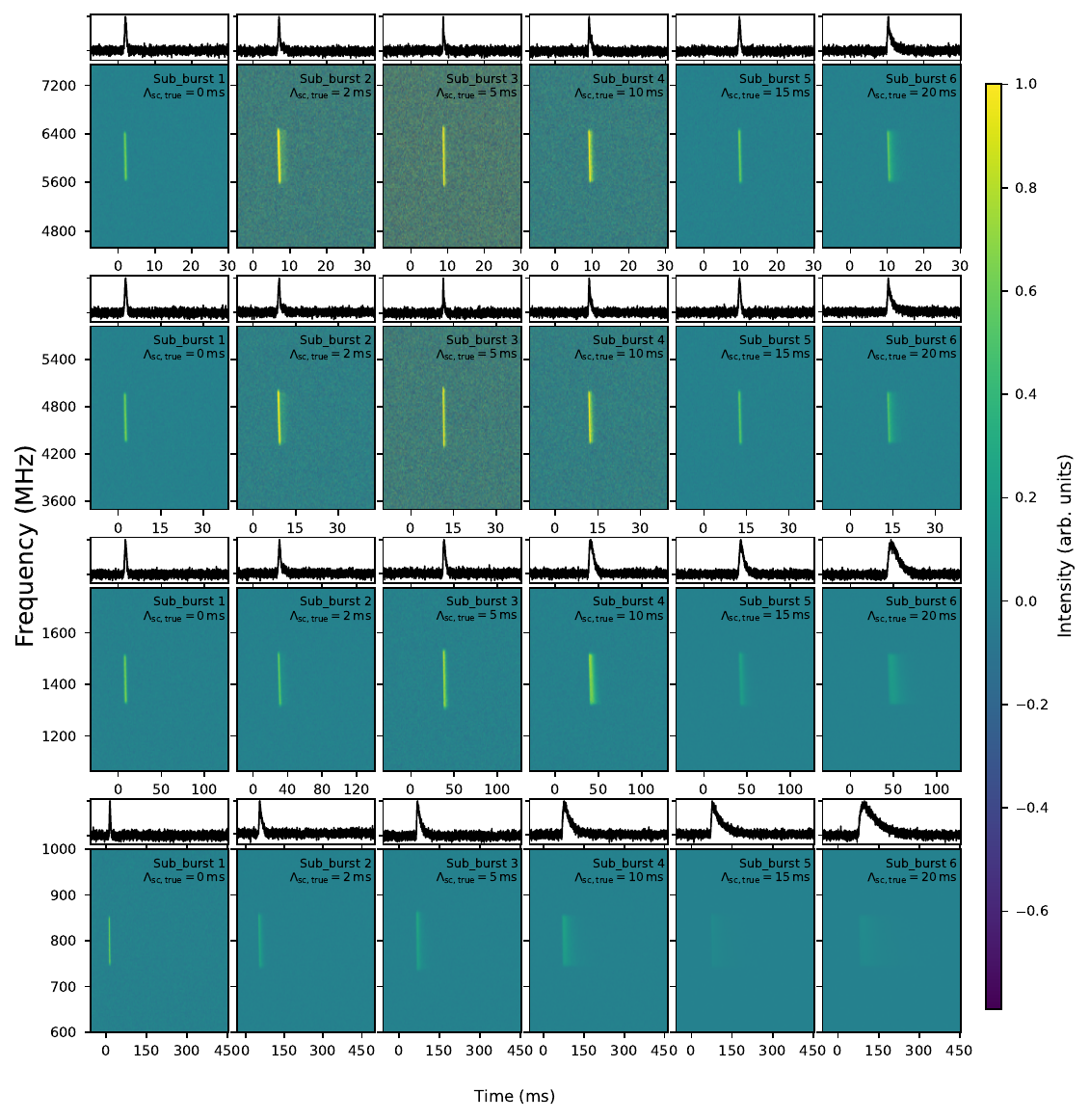}
 \caption{Dynamic spectra of a subset of simulated sub-bursts used in our analysis, shown across multiple observing frequencies and scattering timescales. Each tile shows the frequency-time waterfall of an individual sub-burst, with the normalized frequency-integrated pulse profile (in arbitrary units) plotted above. Rows correspond to our four observing bands (as labeled on the left), centered at 800 MHz, 1420 MHz, 4660 MHz, and 6030 MHz. Each column of dynamic spectra corresponds to a distinct scattering population characterized by a fixed value  $\Lambda_{\mathrm{sc,true}}$, increasing from left to right, as indicated in the upper-right corner of each dynamic spectrum. }
 \label{fig:sub_tile}
\end{figure*}

To illustrate the diversity of spectro-temporal profiles in our simulated FRB population, Figure~\ref{fig:sub_tile} presents six intrinsic superradiant (SR) profiles, each assigned a distinct scattering timescale as indicated in the corresponding dynamic spectra. These profiles are Doppler-shifted to generate emission across four observing frequency bands, producing a representative set of sub-bursts. Residual dispersion is subsequently applied to each burst, enabling an analysis of both propagation effects.

\bibliographystyle{elsarticle-harv} 
\bibliography{refs}






\end{document}